\documentclass[traditabstract]{aa} 

\usepackage{graphicx}
\usepackage{txfonts}

\usepackage{amssymb}
\usepackage{epstopdf}
\usepackage{epsfig}
\usepackage{url}
\usepackage{natbib}
\bibpunct{(}{)}{;}{a}{}{,} 

\begin{document}
   \title{Formation of Sco X-1 induced by anomalous magnetic braking of Ap/Bp stars}
 \titlerunning {Formation of Sco X-1 induced by AMB of Ap/Bp stars}
   \author{Wen-Cong Chen $^{1,2,3}$          }

   \institute{$^1$ School of Physics and Electrical Information, Shangqiu Normal University, Shangqiu 476000, China;\\
 $^2$ Department of Physics, University of Oxford, Oxford OX1 3RH, UK;\\
 $^3$ Argelander-Insitut f\"{u}r Astronomie, Universit\"{a}t Bonn, Auf dem H\"{u}gel 71, 53121 Bonn, Germany;
              \email{chenwc@pku.edu.cn} }

   \date{Received  / Accepted }


  \abstract
{Sco X-1 is the brightest persistent X-ray source in the sky. It is generally believed that Sco X-1 is a low-mass X-ray binary containing a neutron star that accretes from a low-mass donor star where the mass transfer is driven by magnetic braking. However, the mass transfer rate predicted by the standard magnetic braking model is at least one order of magnitude lower than the rate inferred by X-ray luminosity. In this work, we investigate whether this source could have evolved from an intermediate-mass X-ray binary including Ap/Bp stars with a slightly strong magnetic field of 300 - 1000 G. The coupling between the magnetic field and an irradiation-driven wind induced by the X-ray flux from the accretor can yield strong magnetic braking that could give rise to a relatively high mass transfer rate. According to the observed orbital period, the mass transfer rate, the mass ratio, and the donor star spectral type, the progenitor of Sco X-1 should be an intermediate-mass X-ray binary including a  1.6 $-$ 1.8 $\rm M_{\odot}$ Ap/Bp donor star in a 1.3 $-$ 1.5 day orbit. We therefore propose that anomalous magnetic braking of Ap/Bp stars provides an alternative evolutionary channel to some of the luminous X-ray sources.
   }

   \maketitle


\section{Introduction}

As the first discovered persistent extra-solar celestial X-ray source, Sco X-1 plays a milestone role in the X-ray astrophysics field (Giacconi et al. 1962). Interestingly, Sco X-1 is the brightest persistent X-ray source observed in the sky so far (Morrison 1967). Based on the observations by the Very Long Baseline Array, its distance was derived to be $2.8\pm 0.3$ kpc (Bradshaw et al. 1999). Recently, \cite{aasi14} predicted that Sco X-1 is a strong source of gravitational waves.

Sco X-1 has been extensively studied for decades at all energy bands from radio to X-ray, especially at optical and X-ray wavelengths. Based on blue-sensitive plates taken from 1889 to 1974, \cite{gott75} obtained an orbital period of 18.9 hours. Recently, \cite{hyne12} accurately detected the orbital period of this source, which is $P=0.787313\pm0.000015~\rm d$.
A 0.13 mag amplitude modulation in the optical B-band light curve implies that the donor star is heated by X-rays (Augusteijn et al. 1992). Sco X-1 probably is a low-mass X-ray binary (LMXB) where the neutron star is accreting matter from a donor star through an accretion disk that was formed by Roche-lobe overflow (Charles \& Coe 2006). By constraining the radial velocity of the donor star and the primary, \cite{stee02} derived an upper limit of the mass ratio of $q\la0.61$. Adopting the inclination measured by \cite{foma01} in the twin radio lobe observations, they proposed a probable mass ratio $q=0.3$ and a donor star mass 0.42 $\rm M_{\odot}$ (Steeghs \& Casares 2002). Based on a Monte Carlo analysis that included all the previously known orbital parameters, the mass ratio for Sco X-1 was constrained by K-correction to lie between 0.28 and 0.51, and the donor star mass is in the range of 0.28 to 0.70 $\rm M_{\odot}$ (Mata S\'{a}nchez et al. 2015). The deep observations suggest that the donor star has a spectral type later than K4 (Mata S\'{a}nchez et al. 2015), hence the upper limit of the intrinsic effective temperature should be 4800 K (Pavlovskii \& Ivanova 2016).

Sco X-1 has been identified to be a Z-type X-ray source (Hasinger \& van der Klis 1989), and \emph{Rossi X-ray Timing Explorer} observations measured its X-ray luminosity to be $2.3\times 10^{38}~\rm erg\,s^{-1}$ (2 - 20 keV) (Bradshaw et al. 1999) \footnote{The observed X-ray luminosity is expected to increase by a factor of a few when the measured lower limit is 0.2 keV.}. According to its bolometric flux of $3.9\times10^{-7}~\rm erg\,s^{-1}\,cm^{-2}$ (Watt et al. 2008), the bolometric luminosity can be estimated to be $3.6\pm0.8\times 10^{38}~\rm erg\,s^{-1}$. When we assume that the accretor is a standard neutron star ($1.4~\rm M_{\odot}$ and 10 km), the corresponding accretion rate is $3.1\pm0.7 \times10^{-8}~\rm M_{\odot}\,yr^{-1}$. The Eddington accretion rate limited by Thompson-scattering can be written as $\dot{M}_{\rm Edd,TS}=4\pi R_{\rm NS}/(0.2(1+X))$, where $R_{\rm NS}$ is the radius of neutron star, and $X$ is the hydrogen abundance of the accretion material. If the accreted material includes 70 \% hydrogen, $\dot{M}_{\rm Edd,TS}=1.8\times10^{-8}~\rm M_{\odot}\,yr^{-1}$ for a standard neutron star. Therefore, Sco X-1 is expected to experience a super-Eddington accretion.

However, it is still unknown why this LMXB produces such a high mass-transfer rate. When we assume that this source includes a $1.4~\rm M_{\odot}$ neutron star and a $0.4~\rm M_{\odot}$ subgiant companion, the standard magnetic braking given by \cite{rapp83} predicts a mass-transfer rate in the range of 2.6 to $6.0\times10^{-10}~\rm M_{\odot}\,yr^{-1}$ (this depends on the dimensionless parameter $\gamma=0-4$) (Pavlovskii \& Ivanova 2016). It is clear that the theoretical mass-transfer rate is roughly two orders of magnitude lower than the observed value in Sco X-1. \cite{pods02} have noted that most short-period ($\la1~\rm day$) LMXBs have an observed X-ray luminosity higher than expected from theoretical simulation. To solve this problem, \cite{pavl16} proposed a wind-boosted magnetic braking prescription, in which the wind loss in evolved stars is obviously higher than the mass loss in main-sequence stars. Based on this modified prescription, the authors found that the most likely donor-star mass in the zero-age main sequence is 1.4 to $1.5~\rm M_{\odot}$.

Assuming that the donor star evolved from an intermediate-mass Ap/Bp star with an anomalously strong magnetic field (100 -10000 G), \cite{just06} found that an anomalous magnetic braking scenario can form compact black-hole LMXBs. In this work we explore whether the anomalously high X-ray luminosity of Sco X-1 could be interpreted by an anomalous magnetic braking scenario. In Sect. 2 we describe the input physics for the stellar evolution calculation. A numerical calculation result for Sco X-1 is presented in Sect. 3. Finally, we provide a brief discussion and conclusion in Sect. 4.

\section{Stellar evolution code}
We here assumed that the progenitor of Sco X-1 is a binary system containing a neutron star (with a mass of $M_{\rm NS}$) and an intermediate-mass Ap/Bp donor star (with a mass of $M_{\rm d}$)  with an anomalous strong magnetic field. The basic calculation tool is the MESAbinary code in MESA, described in \cite{paxt15}. The evolutionary beginning is a binary system where the donor star at the zero-age main sequence orbits a neutron star on a circular orbit. We took solar chemical compositions ($X=0.7, Y=0.28$, and $Z=0.02$) for the donor star. The neutron star was assumed to be a point mass, and then we calculated the evolution of the donor star and the binary parameters.

When the donor star overflows its Roche lobe in a long-term nuclear evolution, the surface material begins to transfer onto the neutron star at a rate of $\dot{M_{\rm tr}}$, and the binary appears as an intermediate-mass X-ray binary (IMXB). During the accretion of the neutron, the Eddington-accretion rate $\dot{M}_{\rm Edd}$ is taken into account. The Eddington-accretion rate can be written as
\begin{equation}
\dot{M}_{\rm Edd}=3.6\times10^{-8}\left(\frac{M_{\rm NS}}{1.4\rm M_{\odot}}\right)\left(\frac{0.1}{\eta}\right)
\left(\frac{1.7}{1+X}\right)\,\rm M_{\odot}\,yr^{-1},
\end{equation}
where $\eta=GM_{\rm NS}/(R_{\rm NS}c^{2})$ is the energy-conversion
efficiency ($G$ is the gravitational constant, and $c$ is the speed of light).
Therefore, the accretion rate of the neutron star
is $\dot{M}_{\rm NS}={\rm min}(-\dot{M_{\rm tr}},\dot{M}_{\rm Edd})$.
During accretion, the remaining material that exceeds the Eddington accretion rate is thought to be ejected by the radiation pressure, and it forms an isotropic wind in the vicinity of the neutron star.

By the accretion, the gravitational potential energy of the falling material was converted into X-ray radiation. The X-ray luminosity of the neutron star is $L_{\rm X}=GM_{\rm NS}\dot{M}_{\rm NS}/R_{\rm NS}$.  In a relatively compact binary, the strong X-ray radiation of the donor-star surface would induce a stellar wind (Ruderman 1989; Tavani \& London 1993). When we assume that a fraction of $f_{\rm ir}$ the X-ray luminosity received by the donor star can be used to overcome the gravitational potential energy of the wind, the irradiation-driving wind-loss rate is given by
\begin{equation}
\dot{M}_{\rm w}=-f_{\rm ir}L_{\rm X}\frac{R_{\rm d}^{3}}{4 G M_{\rm d} a^{2}},
\end{equation}
where $f_{\rm ir}$ is the irradiation efficiency, and $R_{\rm d}$ and $a$ are the donor-star radius and the orbital separation, respectively. The mass-loss rate of the donor star thus includes the mass-transfer rate and the wind-loss rate, namely $\dot{M}_{\rm d}=\dot{M}_{\rm tr}+\dot{M}_{\rm w}$. In principle, the irradiation process of X-rays has another possible influence, that is, it might alter the effective surface-boundary condition of the donor star, especially the ionization degree of H at the bottom of the irradiated layer (Podsiadlowski 1991). Considering the effect of X-ray illumination of the donor star in a close binary on its internal structure, Hameury et al. (1993) found  episodic irradiation-induced mass-transfer cycles during the evolution of LMXBs. However, the X-ray illumination model proposed by these authors cannot account for a system with orbital periods longer than 10 hr, which is obviously shorter than the orbital period of Sco X-1. We therefore ignored the X-ray illumination effect in the calculation.

 We assumed that the magnetic field of the donor star obeys a dipolar distribution, and declines with distance as $B(r)=B_{\rm s}R_{\rm NS}^{3}/r^{3}$ (where $B_{\rm s}$ is the surface magnetic field of the donor star). The winds are thought to be ejected from the surface of the donor star at the escape speed (Justham et al. 2006). During the wind outflows, the strong magnetic field of Ap/Bp stars can control the charge particles inside the magnetospheric radius. Similar to \cite{ivan06}, the wind-loss rate was assumed to be constant during outflow. By balancing between the magnetic pressure and ram pressure, the magnetospheric radius can be derived as (Justham et al. 2006)
\begin{equation}
r_{\rm m}=(GM_{\rm d})^{-1/8}B_{\rm s}^{1/2}R_{\rm
d}^{13/8}\dot{M}_{\rm w}^{-1/4}.
\end{equation}
 Assuming that the outflow
material is tied in the magnetic field lines to corotate with the stars and that it escapes from the influence of the magnetic field at $r_{\rm m}$, the loss rate of the spin angular momentum is $\dot{J}_{\rm s}=2\pi\dot{M}_{\rm w}r^{2}_{\rm m}/P$, where $P$ is the orbital period. With the spin-down of the evolved star, the tidal interaction between the two components would continuously spin
the star back up into corotation with the orbital rotation (Patterson 1984). As a result, the magnetic braking of the donor star is an efficient mechanism that indirectly carries away the orbital angular momentum. Therefore the loss rate of angular-momentum
produced by magnetic braking is (Justham et al. 2006)
\begin{equation}
\dot{J}_{\rm mb}=-B_{\rm s}\left(\frac{\dot{M}_{\rm w}^{2}R_{\rm d}^{13}}{GM_{\rm d}}\right)^{1/4}\frac{2\pi}{P}.
\end{equation}

\begin{figure}
\centering
\includegraphics[angle=0,width=10cm]{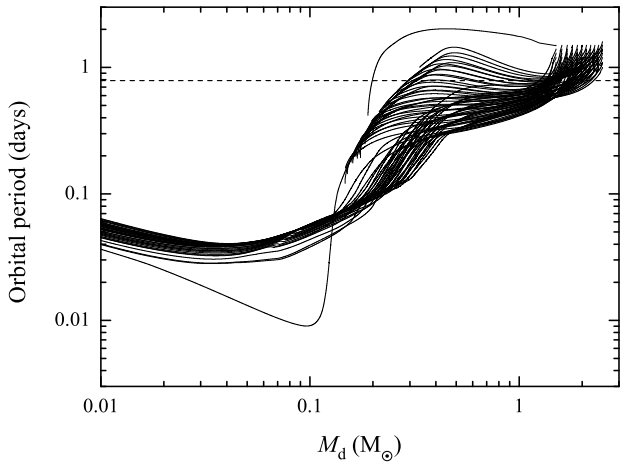}
\caption{Evolutionary tracks of IMXBs
with different initial donor-star masses and initial orbital periods in the orbital period vs. donor-star mass plane when the surface magnetic field of the donor star $B_{\rm s}=1000~\rm G$. The horizontal line corresponds to the observed orbital period of Sco X-1. } \label{Fig1}
\end{figure}

In the numerical calculation, orbital angular momentum loss plays a vital role for the evolution of IMXBs. We considered the following three mechanisms: (1) gravitational-wave radiation; (2)
anomalous magnetic braking (see also Equation 4), which would work when the mass transfer begins and X-rays could induce a wind from the donor star; (3) mass loss: the
mass loss from the vicinity of the pulsar is
assumed to carry away the specific orbital angular
momentum of the pulsar, while the donor star winds carry away the specific orbital angular momentum of the donor star.

\begin{figure}
\centering
\includegraphics[angle=0,width=10cm]{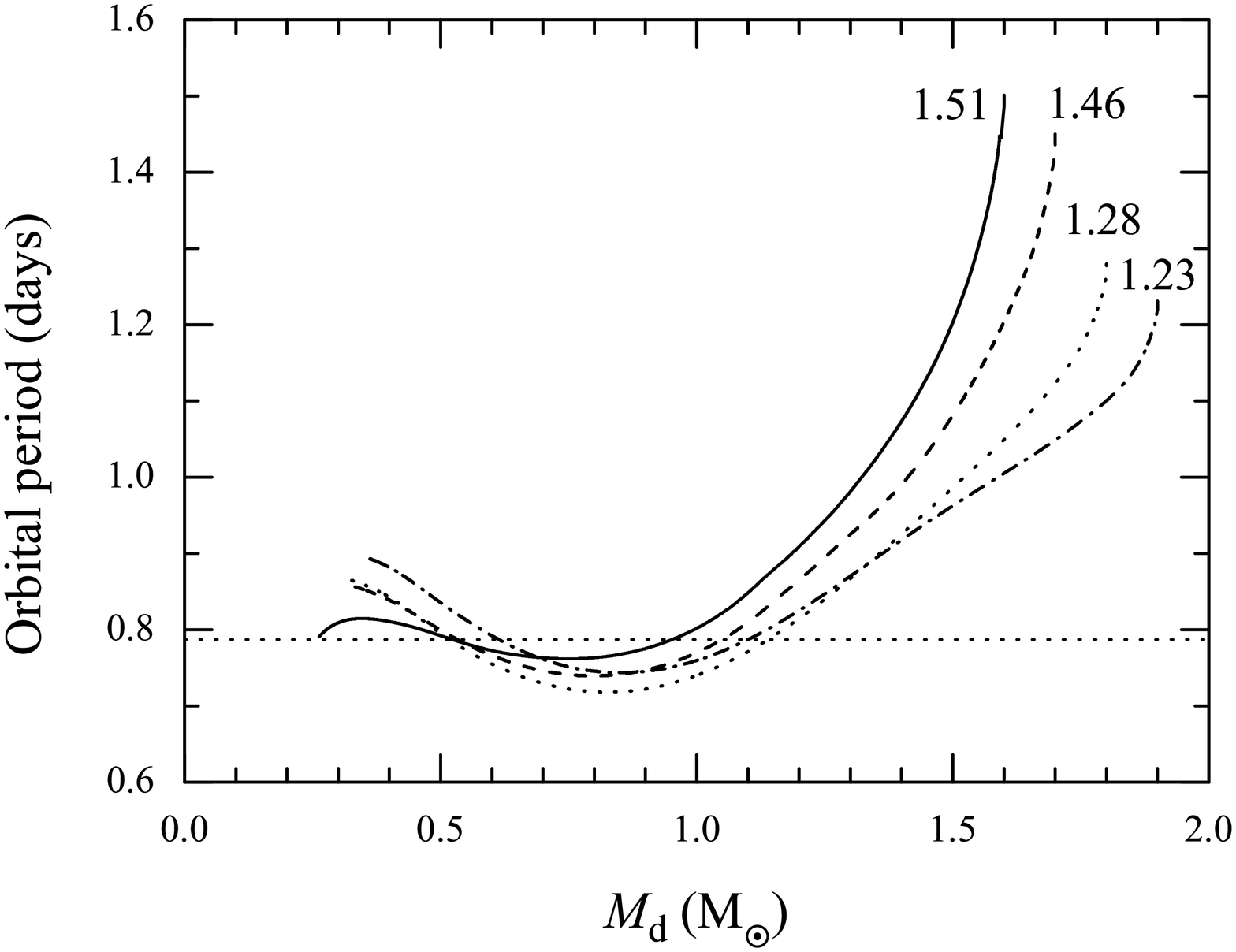}
\includegraphics[angle=0,width=10cm]{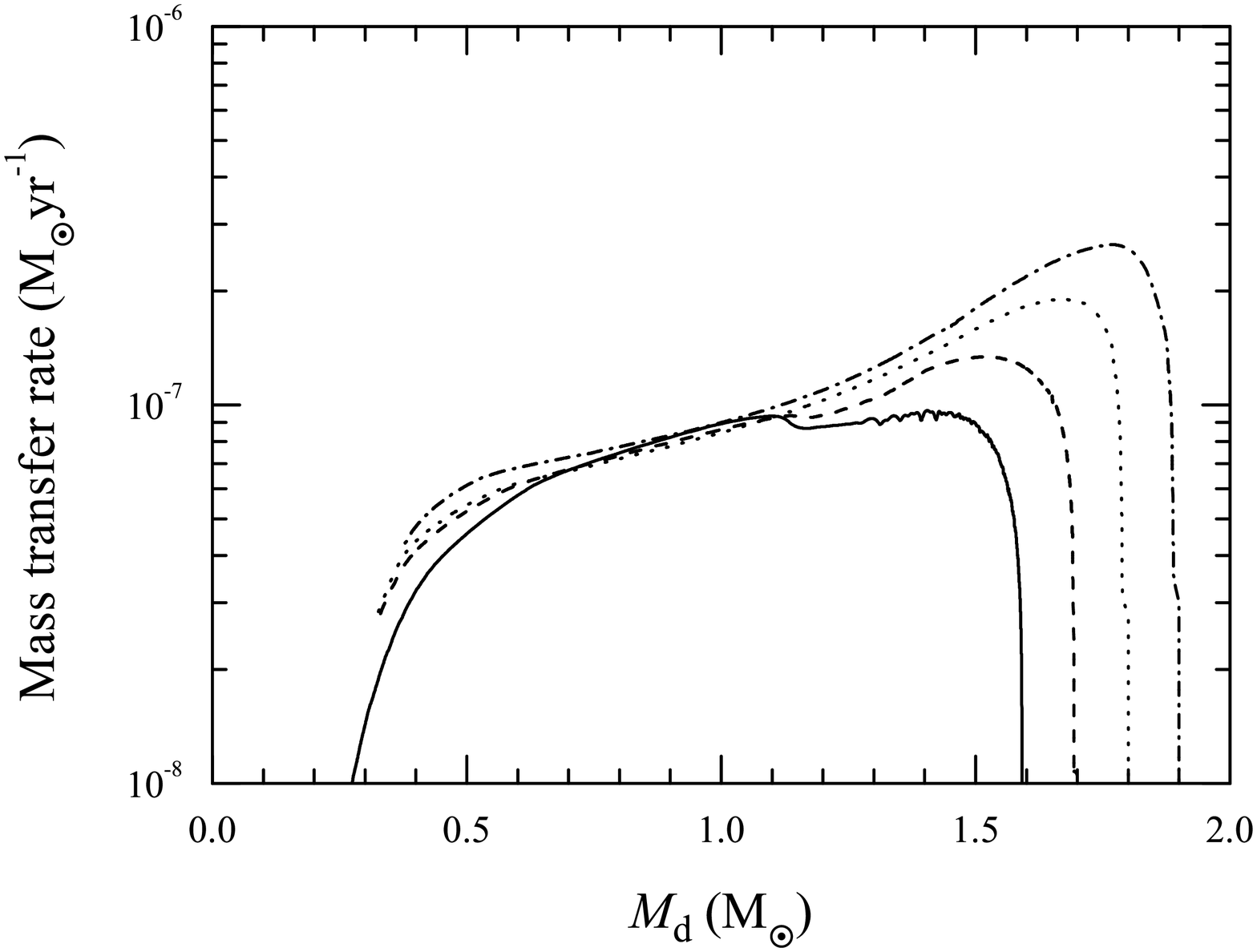}
\caption{Evolutionary tracks of IMXB
with different initial donor-star masses and initial orbital periods in the orbital period vs. donor-star mass (top panel) and
the mass-transfer rate vs. donor-star mass (bottom panel) diagram when the surface magnetic field of the donor star $B_{\rm s}=300~\rm G$. The solid, dashed, dotted, and dash-dotted curves
represent a donor-star mass of 1.6, 1.7, 1.8, and 1.9 $\rm M_{\odot}$, respectively. Numbers inside the curves denote the initial orbital periods in units of days. The horizontal line in the top panel corresponds to the observed orbital period of Sco X-1. } \label{Fig1}
\end{figure}

\section{Results}

\begin{table*}
\centering
\begin{minipage}{150mm}
\caption{Calculated relevant parameters of IMXBs with different initial orbital periods and donor-star masses. The meaning of the columns is as follows: the surface magnetic field, the initial donor-star mass, the initial orbital period, and the orbital period, the age, the donor-star mass, the neutron-star mass, the mass ratio,  the mass-transfer
rate, the wind-loss rate, the donor-star radius, the effective temperature, and the center He abundance when when the orbital period can fit the observed values of Sco X-1 in the period-increasing phase.}
\begin{tabular}{lllllllllllll}
\hline  \hline\noalign{\smallskip}
$B_{\rm s}$ & $M_{\rm d,i}$ & $P_{\rm orb,i}$  & $P_{\rm orb}$ & Age     & $M_{\rm d}$ & $M_{\rm NS}$ & $q$& $\dot{M}_{\rm tr}$ & $\dot{M}_{\rm wind}$ & $R_{\rm d}$ & $T_{\rm eff}$ & $Y_{\rm c}$ \\
(G) & $(\rm M_{\odot})$  &(d)  & (d) &  (Gyr) & ($\rm M_{\odot}$) & ($\rm M_{\odot}$) &  & ($\rm M_{\odot}\,yr^{-1}$) & ($\rm M_{\odot}\,yr^{-1}$) & ($\rm R_{\odot}$)& (K) & \\
 \hline\noalign{\smallskip}
300&1.6& 1.51 &0.787& 1.83 &0.488& 1.62 & 0.30 & $2.5\times10^{-8}$ & $1.0\times10^{-7}$ &1.29 & 4767 &0.98\\
   &1.7& 1.46 &0.787& 1.50 &0.521& 1.61 & 0.32 & $2.8\times10^{-8}$ & $1.0\times10^{-7}$ &1.32 & 4786 &0.98\\
   &1.8& 1.28 &0.787& 1.24 &0.511& 1.63 & 0.31 & $2.9\times10^{-8}$ & $1.0\times10^{-7}$ &1.32 & 4786 &0.98\\
   &1.9& 1.23 &0.787& 1.05 &0.603& 1.60 & 0.38 & $3.9\times10^{-8}$ & $1.0\times10^{-7}$ &1.39 & 4897 &0.98\\
\hline
1000&1.8&1.81 &0.787& 1.28 &0.509& 1.49 & 0.34 & $1.0\times10^{-7}$ & $1.0\times10^{-7}$ &1.32 & 4862 &0.98 \\
    &1.9&1.58 &0.787& 1.08 &0.459& 1.50 & 0.31 & $0.9\times10^{-7}$ & $1.0\times10^{-7}$ &1.27 & 4894 &0.98 \\
    &2.0&1.40 &0.787& 0.92 &0.449& 1.50 & 0.30 & $0.9\times10^{-7}$ & $1.0\times10^{-7}$ &1.26 & 4953 &0.98 \\
    &2.1&1.33 &0.787& 0.80 &0.496& 1.49 & 0.33 & $1.0\times10^{-7}$ & $1.0\times10^{-7}$ &1.30 & 4996 &0.98 \\
\noalign{\smallskip}\hline
\end{tabular}
\end{minipage}
\end{table*}

Our anomalous magnetic braking scenario must be based on an intermediate-mass donor star, so the donor-star masses are taken to be greater than $1.5~\rm M_{\odot}$. When the mass ratio of an IMXB is in the range of 0.3 $-$ 0.4 when its orbital period is 0.787 d, it represents the candidate progenitor of Sco X-1. In our standard model, a typical neutron-star mass of $1.4~\rm M_{\odot}$ and a constant radius of $10~\rm km$ were assumed. According to Eqs. 2 and 4, $\dot{J}_{\rm mb}\propto B_{\rm s}f_{\rm ir}^{1/2}$, which means that a degenerate relation exists between $B_{\rm s}$ and $f_{\rm ir}$. We changed the surface magnetic field $B_{\rm s}$, while the irradiation efficiency was assumed to be a fixed value ($f_{\rm ir}=10^{-3}$).

To understand the progenitor properties of Sco X-1, we
calculated the evolution of 66 IMXBs in the $P_{\rm orb,i}-M_{\rm d,i}$ plane, which was divided into $11\times6$ discrete
grids ($M_{\rm d,i}=1.5-2.5~\rm M_{\odot},P_{\rm orb,i}=1.0-1.5~\rm d$) (see also Figure 1). All evolutionary tracks can be divided into two parts: (i) some IMXBs are expected to evolve into ultra-compact X-ray binaries if their initial orbital-periods are greater than the so-called bifurcation period, and (ii) the remaining IMXBs evolve into detached binaries consisting of a neutron star and a white dwarf. To fit the observed parameters of Sco X-1, the donor-star mass needs to be $0.45-0.6~\rm M_{\odot}$ when $P_{\rm orb}=0.787~\rm d$. Sco X-1 belongs to the second part, and only two IMXBs can evolve into the Sco X-1-like systems. It therefore seems that the initial parameter space of Sco X-1 is very narrow.

Figure 2 summarizes the evolution of the orbital period and the mass-transfer rate of four IMXBs when the surface magnetic field of the donor star $B_{\rm s}=300$ G. In the first stage, the orbital period of the binary continuously decreases as a result of thermal timescale mass transfer ($M_{\rm d}>M_{\rm NS}$) and strong magnetic braking ($M_{\rm d}<M_{\rm NS}$). Our simulation shows that the mass transfer needs to be super-Eddington and to give rise to a high X-ray luminosity and strong winds from the donor star. However, the mass transfer would induce the orbit to broaden after the material was transferred from the less massive donor star to the more massive neutron star. When the effects produced by two cases (angular momentum loss and mass transfer) balance out, the IMXB would reach a minimum orbital period. When the donor-star masses are lower than $0.4~\rm M_{\odot}$, the mass-transfer rate sharply declines and results in a weak winds. As a result, the orbital period begins to decrease again.

The evolutionary tracks of donor stars in H-R diagram of four cases in Figure 2 are illustrated in Figure 3. It is clear that the center H is already exhausted before the mass transfer begins. Therefore, it seems that Sco X-1 experiences a Case B mass transfer. Subsequently, the mass transfer and wind loss would cause the luminosity and effective temperature of the donor stars to decrease.

\begin{figure}
\centering
\includegraphics[angle=0,width=10cm]{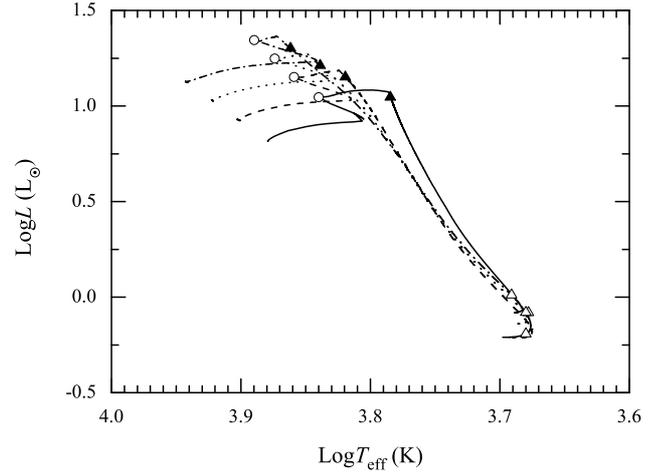}
\caption{Evolutionary tracks of IMXBs
with different initial donor-star masses and initial orbital periods in the H-R diagram when the surface magnetic field of the donor star $B_{\rm s}=300~\rm G$. The meaning of the four curves is same as in Figure 1. The open circles, filled triangles, and open triangles represent the moment that the center H is exhausted, the beginning of mass transfer, and the moment when the orbital periods are 0.787 days. } \label{Fig1}
\end{figure}

Table 1 presents the detailed binary and stellar parameters when the calculated orbital periods can fit the observed values of Sco X-1 in the period-increasing phase. The current mass-transfer rates are  2.5, 2.8, 2.9, and $3.9\times10^{-8}~\rm M_{\odot}\,yr^{-1}$ for a donor star with a mass from 1.6 to $1.9~\rm M_{\odot}$. Our simulated mass-transfer rates are approximately consistent with the inferred value in observations. At $P=0.787$ d, all donor stars exhaust their central hydrogen and are experiencing a Case B mass transfer. Because the neutron stars in four cases experience a super-Eddington accretion at the current period, the Eddington luminosity produces the same wind-loss rate from the donor star according to Equation (2). The current effective temperatures of the donor stars increase with their initial masses. A donor star with an initial mass of $1.9~\rm M_{\odot}$ has an effective temperature of 4897 K, which obviously exceeds the observed upper limit (4800 K). Therefore, it seems that the progenitor donor-star mass is unlikely to be $\ga 1.9~\rm M_{\odot}$ when the donor star possesses a surface magnetic field of 300 G. Our simulation shows that the current donor star in Sco X-1 probably has a mass of $0.48-0.52~\rm M_{\odot}$ and a radius of $1.3~\rm R_{\odot}$.

\begin{figure}
\centering
\includegraphics[angle=0,width=10cm]{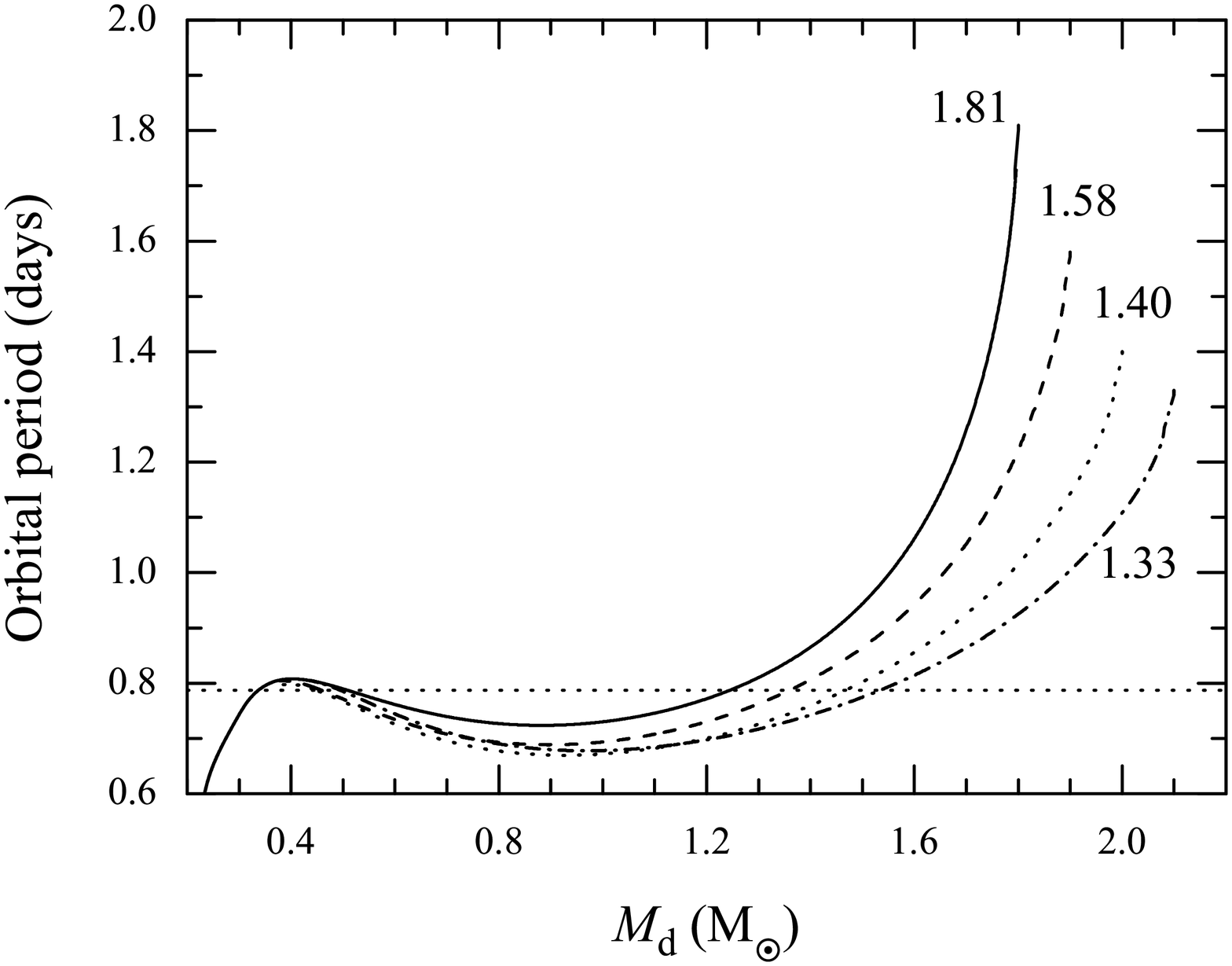}
\includegraphics[angle=0,width=10cm]{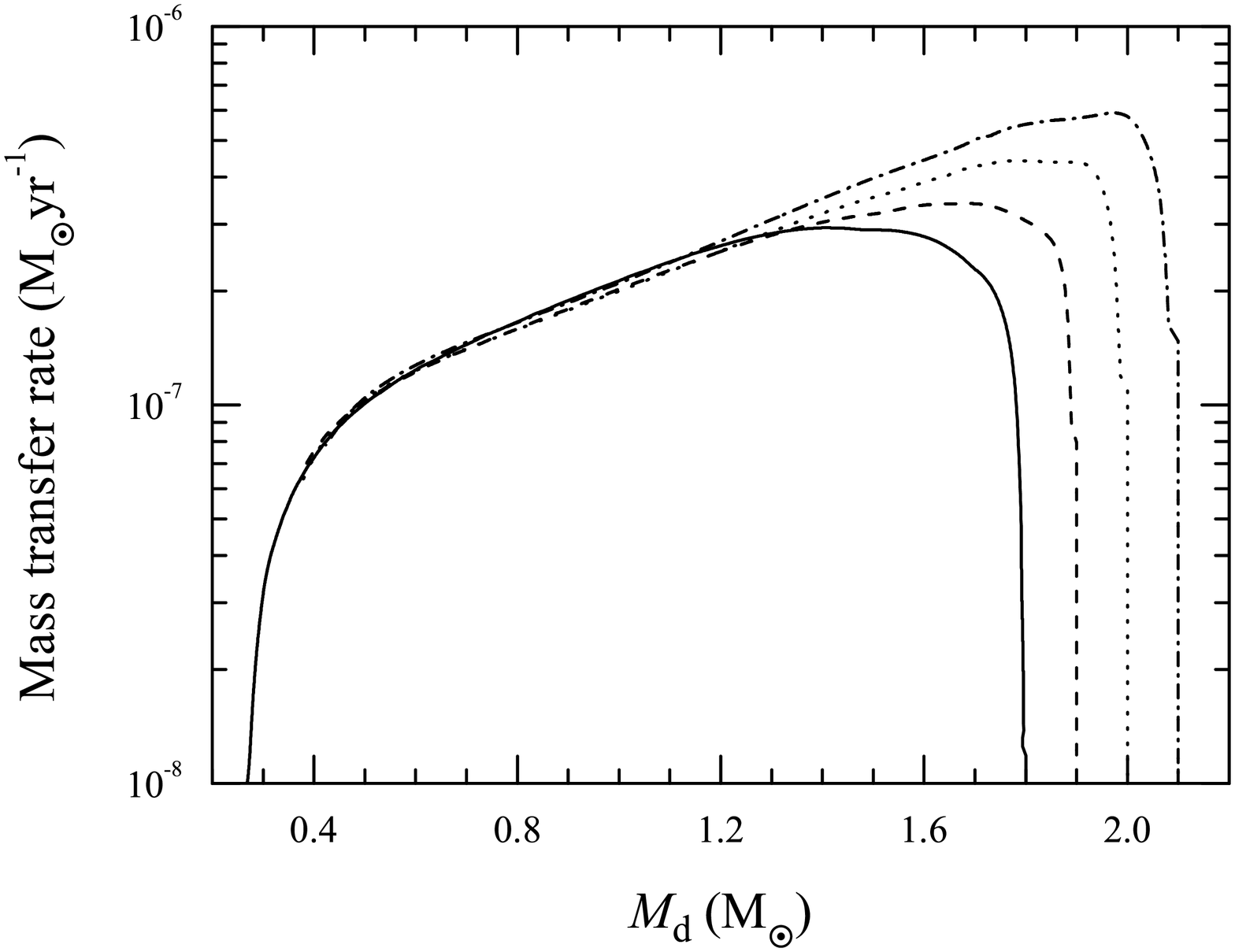}
\caption{Same as in Fig. 2, but for $B_{\rm s}=1000~\rm G$. The solid, dashed, dotted, and dash-dotted curves
represent donor-star masses of 1.8, 1.9, 2.0, and 2.1 $\rm M_{\odot}$, respectively. } \label{Fig1}
\end{figure}

When $B_{\rm s}=1000$ G, the evolutionary tracks of the orbital period and the mass-transfer rate of IMXBs are plotted in Figure 4. Because the magnetic fields increase by a factor of 10/3, the angular momentum loss rate is roughly enhanced by a similar factor (we cannot expect that strictly $\dot{J}_{\rm }\propto B_{\rm s}$ since the wind-loss rate would accordingly increase). When $P_{\rm orb}=0.787~\rm d$, a relatively strong magnetic braking has driven a very high mass-transfer rate ($\sim10^{-7}~\rm M_{\odot}\,yr^{-1}$ ), which is obviously higher than the case with a weak magnetic field (300 G). Meanwhile, all donor stars have a current effective temperature greater than 4800 K (see also Table 1), which was thought to be the maximum effective temperature of the donor star in Sco X-1. The possibility that this source evolved from an IMXB including a Ap/Bp star with a strong surface magnetic field can therefore be ruled out. Because the initial orbital periods are longer than the so-called bifurcation periods (see also Figure 4 of \cite{chen16}, two figures have the same magnetic-braking strength), it is impossible for these systems to evolve into a compact orbit. Figure 5 shows the evolution of six key parameters for an LMXB with a $2.0\rm M_{\odot}$ donor star and an initial orbital period of 1.4 d, which can form Sco X-1. At the age of 923 Myr, the system became a Sco X-1-like system (see also Table 1). It is clear that the mass transfer always occurs at a high rate $\ga 10^{-7}~\rm M_{\odot}\,yr^{-1}$ until the IMXB evolves into a Sco X-1-like system. During the nuclear evolution of the donor star, it developed a convective envelope with a mass fraction of 5\%. Adopting the equation given by Hurley et al (2002) for a convective envelope, the tidal synchronization timescales of the donor star are $\sim100 - 300$ yr. Since the mass-transfer timescale (about 10 Myr) is five orders of magnitude greater than the tidal synchronization timescales, the tidal interaction can keep the donor star spin synchronous with the orbital motion.

\begin{table*}
\centering
\begin{minipage}{150mm}
\caption{Similar to Table 1, but for an initial neutron-star mass of $1.2\rm M_{\odot}$.}
\begin{tabular}{lllllllllllll}
\hline  \hline\noalign{\smallskip}
$B_{\rm s}$ & $M_{\rm d,i}$ & $P_{\rm orb,i}$  & $P_{\rm orb}$ & Age     & $M_{\rm d}$ & $M_{\rm NS}$ & $q$& $\dot{M}_{\rm tr}$ & $\dot{M}_{\rm wind}$ & $R_{\rm d}$ & $T_{\rm eff}$ & $Y_{\rm c}$ \\
(G) & $(\rm M_{\odot})$  &(d)  & (d) &  (Gyr) & ($\rm M_{\odot}$) & ($\rm M_{\odot}$) &  & ($\rm M_{\odot}\,yr^{-1}$) & ($\rm M_{\odot}\,yr^{-1}$) & ($\rm R_{\odot}$)& (K) & \\
 \hline\noalign{\smallskip}
300&1.6& 1.51 &0.787& 1.83 &0.542& 1.38 & 0.39 & $3.4\times10^{-8}$ & $1.0\times10^{-7}$ &1.35 & 4805 &0.98\\
   &1.7& 1.46 &0.787& 1.50 &0.460& 1.42 & 0.32 & $2.9\times10^{-8}$ & $1.0\times10^{-7}$ &1.27 & 4770 &0.98\\
   &1.8& 1.28 &0.787& 1.24 &0.474& 1.42 & 0.33 & $2.9\times10^{-8}$ & $1.0\times10^{-7}$ &1.36 & 4786 &0.98\\
   &1.9& 1.21 &0.787& 1.05 &0.521& 1.40 & 0.30 & $3.3\times10^{-8}$ & $1.0\times10^{-7}$ &1.35 & 4841 &0.98\\
\noalign{\smallskip}\hline
\end{tabular}
\end{minipage}
\end{table*}

Similar to \cite{pavl16}, we also investigated the influence of the initial mass of the neutron star on the evolutionary results. In Table 2 we list the calculated results when the mass of the neutron star is 1.2 $\rm M_{\odot}$. Our simulation indicates that the current effective temperatures of the donor star are almost not influenced by the initial neutron-star mass, which is in contrast with the results reported by \cite{pavl16}. The main reason for this difference may be the fixed mass ratio adopted in \cite{pavl16}. In this case, a low mass of the neutron star would naturally
result in a low donor-star mass and a low effective temperature.

\section{Discussion and summary}
Based on the anomalous magnetic-braking model proposed by \cite{just06}, we have investigated whether Sco X-1 could have originated from an IMXB including a intermediate-mass Ap/Bp star. Our simulated results for IMXBs with a 1.6 $-$ 1.8 $\rm M_{\odot}$ donor star can account for the observed orbital period, the possible donor-star mass, the mass-transfer rate, and the effective temperature of Sco X-1 if the Ap/Bp donor star has a surface magnetic field of 300 G. All cases when $B_{\rm s}= 1000$ G were able to produce a mass-transfer rate higher than the observed value, however, the current effective temperatures of donor stars are obviously higher than 4800 K. Therefore, the progenitor of Sco X-1 could be an IMXB consisting of a neutron star and a 1.6 $-$ 1.8 $\rm M_{\odot}$ Ap/Bp star with a magnetic field of $\sim300~\rm G$. Furthermore, our simulation indicates that Sco X-1 would evolve into a low-mass binary pulsar including a white dwarf.

Our calculation confirms that Sco X-1 is experiencing super-Eddington accretion. For a standard neutron star, the accreting-hydrogen fraction of 70\% and opacities provided by Thompson scattering would yield an Eddington luminosity $\dot{L}_{\rm Edd,TS}=2.1\times10^{38}~\rm erg\,s^{-1}$ (Pavlovskii \& Ivanova 2016). However, the observed X-ray luminosity of this source in the 2$-$20 keV range is $2.3\times10^{38}~\rm erg\,s^{-1}$ (Bradshaw et al.
1999), which is higher than the Eddington luminosity by a factor of 10\%. In our calculation, the mass of the neutron star increases by a fraction 10\% of its initial mass, which is consistent with the increasing fraction of the X-ray luminosity. Therefore, it seems that this source still obeys the Eddington accretion theory. All cases in our simulation have a mass-transfer rate higher than the Eddington accretion rate ($\sim2.0\times10^{-8}~\rm M_{\odot}\,yr^{-1}$) of a $\sim1.6~\rm M_{\odot}$ neutron star. This is consistent with evidence that Sco X-1 is a microquasar and has a jet that originates from the accreting neutron star (Mirabel \& Rodr\'{i}guez 1999).

\begin{figure*}
\begin{centering}
\includegraphics[angle=0,width=16cm]{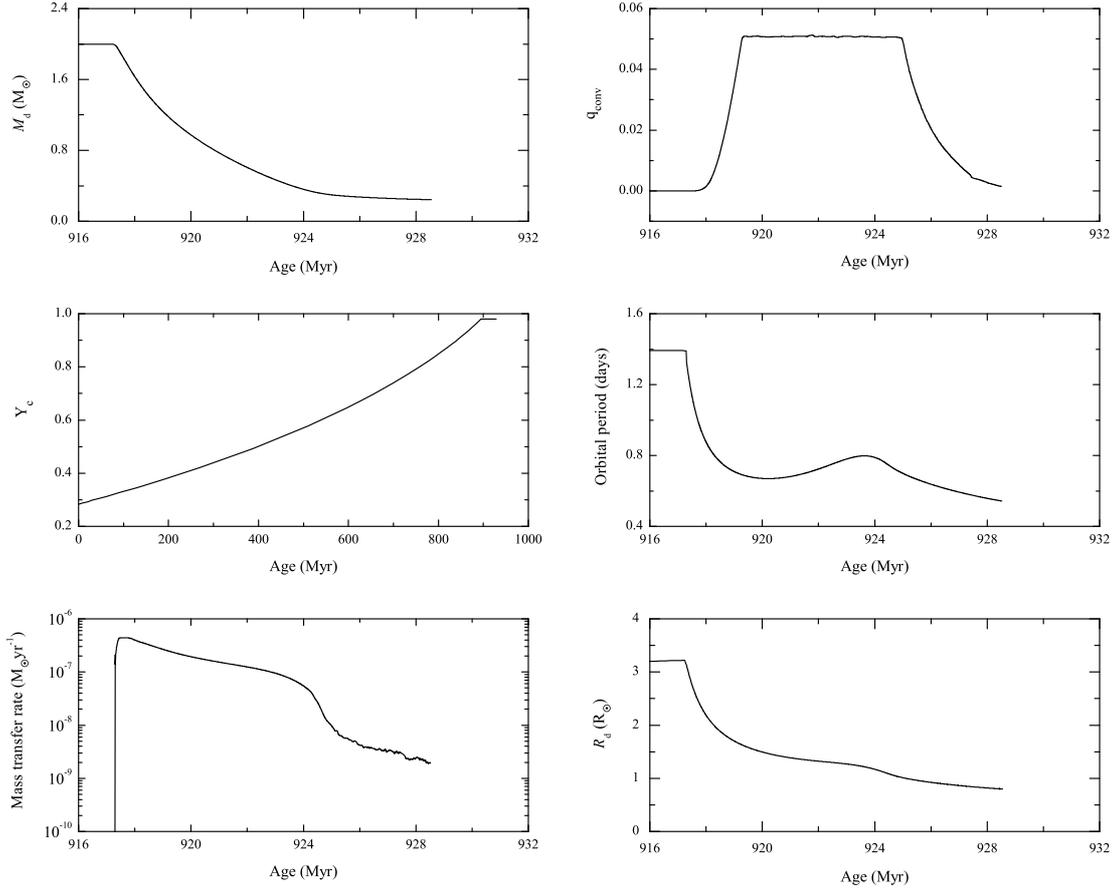}
\par\end{centering}
\caption{Evolution of six key parameters as a function of the donor-star age from the zero-age main sequence for an IMXB with a donor-star mass of $2~\rm M_{\odot}$ and an initial period of 1.4 days. Top left panel: the donor-star mass. Middle left panel: the center He abundance. Bottom left panel: the mass-transfer rate. Top right panel: the mass fraction of the convective envelope. Middle right panel: the orbital period. Bottom right panel: the donor-star radius. } \label{Fig1}
\end{figure*}

The standard magnetic braking can successfully explain the period gap in cataclysmic variables (Spruit \& Ritter 1983). However, some new observations challenge the standard magnetic braking description. Some works have therefore presented modified magnetic braking scenarios. \cite{just06} proposed an anomalous magnetic braking model to account for the formation of compact black-hole LMXBs. In this work, our simulated results considered the anomalous magnetic braking model and agree well with the observed parameters of Sco X-1. \cite{ivan06} proposed that pre-main-sequence donors would induce a stronger magnetic braking and give rise to the formation short-period black-hole LMXBs.

At present, theory and observation for the evolution of LMXBs still disagree. First, a large discrepancy exists for known LMXBs between the inferred mass-transfer rate in observations and the calculated rate based on the standard magnetic braking model (Podsiadlowski et al. 2002). Some short-period LMXBs like Sco X-1 have been reported to have an accretion rate at least one order of magnitude higher than the rate predicted by theoretical simulations. This problem may also be related to the birthrate of millisecond pulsars in the Galactic disk (Kulkarni \& Narayan 1988; Johnston
\& Bailes 1991) and in globular clusters (Fruchter \& Goss 1990; Kulkarni et al. 1990), in which the birthrate is 1$-$2 orders of magnitude higher than that of LMXBs. Second, the three black-hole LMXBs XTE J1118,  A0620-00, and Nova Muscae 1991 have recently been reported to have a very fast orbital decay (Gonz\'{a}lez Hern\'{a}ndez et al. 2012, 2014, 2017). Neither gravitational radiation nor standard magnetic braking can produce such a very high orbital-period derivative.

The birthrate problem probably originates from an overestimate of the LMXBs lifetimes ($\sim 5\times10^{9}~\rm yr$, Podsiadlowski et al. 2002). If LMXBs have a mean mass-transfer rate of $\sim10^{-8}~\rm M_{\odot}\,yr^{-1}$, their lifetime is $\sim 10^{8}~\rm yr$, and the birthrate problem would vanish. Therefore, an angular-momentum-loss mechanism much stronger than the standard magnetic braking is required during the evolution of LMXBs. On the other hand, an efficient angular-momentum-loss mechanism is probably also responsible for the fast orbital decay detected in three black-hole LMXBs. However, only 5\% A/B stars
have anomalously strong magnetic fields (Landstreet 1982; Shorlin et al. 2002), hence anomalous magnetic braking cannot fully solve the birthrate problem. As a possibility, it provides an alternative evolutionary channel toward some luminous LMXBs including Sco X-1 and ultra-compact X-ray binaries (Chen \& Podsiadlowski 2016). If the donor stars have a very strong magnetic field and a relatively high wind-driving efficiency, anomalously high orbital-period derivatives of black-hole LMXBs could be produced by anomalous magnetic braking (Gonz\'{a}lez Hern\'{a}ndez et al. 2017). In addition, circumbinary disks surrounding the binary may be an alternative mechanism as well (Chen \& Li 2015; Chen \& Podsiadlowski 2017 ).

\begin{acknowledgements}
We are grateful to the anonymous referee for very helpful
and useful suggestions. This work was partly supported by the National Natural Science Foundation of China (under grant number 11573016), the Program for Innovative Research Team (in Science and Technology) at the University of Henan Province, and the China Scholarship
Council.
\end{acknowledgements}

\bibliographystyle{aa}
\bibliography{biblio_NH2CHO}

\end{document}